\documentclass{llncs}
\usepackage{times}
\usepackage{url}
\usepackage{fancyvrb}
\usepackage{color}
\usepackage{ifthen}
\usepackage{calc}
\usepackage[pdftex]{graphicx}
\DeclareGraphicsExtensions{.pdf,.jpg,.png}
\graphicspath{{figs/}{./}}
\begin{document}

\title{SWISH: SWI-Prolog for Sharing}
\pagestyle{empty}

\author{Jan Wielemaker\inst{1} and
	Torbj\"orn Lager\inst{2} and
	Fabrizio Riguzzi\inst{3}}
\institute{Web and Media group,
	   VU University Amsterdam, The Netherlands, \\
	   \email{J.Wielemaker@vu.nl}
	   \and
	   Department of Philosophy, Linguistics and Theory of Science,
	   University of Gothenburg, Sweden \\
	   \email{Torbjorn.Lager@ling.gu.se}
	   \and
	   Dipartimento di Matematica e Informatica,
	   University of Ferrara, Italy \\
	   \email{Fabrizio.Riguzzi@unife.it}
	  }

\pagestyle{plain}
\setcounter{page}{99}

\maketitle
\bgroup
\newcommand{\TBD}[1]{\textbf{#1}}
\input{swipl.sty}

\begin{abstract}
Recently, we see a new type of interfaces for programmers based on web
technology. For example, JSFiddle, IPython Notebook and R-studio. Web
technology enables cloud-based solutions, embedding in tutorial web
pages, attractive rendering of results, web-scale cooperative
development, etc. This article describes SWISH, a web front-end for
Prolog. A public website exposes SWI-Prolog using SWISH, which is used
to run small Prolog programs for demonstration, experimentation and
education. We connected SWISH to the ClioPatria semantic web toolkit,
where it allows for collaborative development of programs and queries
related to a dataset as well as performing maintenance tasks on the
running server and we embedded SWISH in the Learn Prolog Now! online
Prolog book.
\end{abstract}


\section{Introduction}
\label{sec:intro}

Web technology has emerged to a state where it becomes useable for
implementing programming development environments. All major modern
browsers now implement HTML5 and JavaScript and there are mature
components available such as the
CodeMirror\footnote{\url{https://codemirror.net/}} and
ACE\footnote{\url{http://ace.c9.io}} code editors, the
Bootstrap\footnote{\url{http://getbootstrap.com/}} framework for styling
and UI widgets and vizualization libraries such as
D3.js.\footnote{\url{http://d3js.org/}} Using web technology rather than
traditional GUI based technology such as
Eclipse,\footnote{\url{https://eclipse.org/}} Microsoft Visual
Studio,\footnote{\url{https://www.visualstudio.com/}}
XEmacs,\footnote{\url{http://www.xemacs.org/}} etc.\ has various
advantages. Being network transparent, it allows for controlling cloud
hosted applications as well as Prolog processes running on headless
devices. Web technology provides a great infrastructure for
\textit{mashups}, pages that integrate material from several sources.
For example, embedding Prolog in tutorial pages or embedding Prolog
queries that can be modified and re-evaluated in documents that describe
data collections.

With SWISH (\textbf{SWI}-Prolog for \textbf{Sh}aring), we provide this
technology for (SWI-)Prolog. SWISH consists of JavaScript client
(browser) code and a number of Prolog libraries that realise the server
as a Prolog application. The client code consists of a series of
jQuery\footnote{\url{https://jquery.com/}} plugins that deal with
editing source code, managing a shared source repository, entering
queries and rendering answers produced by Prolog. The server-side
libraries serve the overall web application, implement the source store
and support the editor with predicate documentation, templates,
cross-reference results, etc. For executing Prolog queries, SWISH relies
on \textit{Pengines} (Prolog engines,
\cite{DBLP:journals/tplp/LagerW14}). A pengine is a Prolog engine that
can be controlled similarly to Prolog running in a terminal using HTTP
requests. The SWISH infrastructure was originally developed as a Prolog
version of JSFiddle. It was later reimplemented as a modular jQuery
based infrastructure aiming at collaborative exploration of data hosted
on a SQL or SPARQL server. This use-case is described in
\secref{cliopatria}.

This article is organised as follows. \Secref{related} describes related
work, which in our case are the systems that have inspired us.
\Secref{application} describes the architecture and components of SWISH.
In \secref{applications} we describe four applications of the current
system.  We conclude with future work and conclusions.

\section{Related work}
\label{sec:related}

We are not aware of other initiatives that aim at developing a rich
web-based development environment for Prolog. We do not compare SWISH
with traditional editor or GUI based development environments for Prolog
because web-based environments provide new opportunities and pose new
challenges. Instead, we discuss three applications that have served as
inspiration for SWISH: JSFiddle,\footnote{\url{https://jsfiddle.net/}}
R-Studio\footnote{\url{http://www.rstudio.com/}} and IPython
Notebook.\footnote{\url{http://ipython.org/notebook.html}}

\begin{itemize}
\item
As stated, the initial inspiration for SWISH was JSFiddle. Unlike
JSFiddle though, Prolog is executed on the server rather than in the
browser.

\item
R-Studio \cite{gandrud2013reproducible} is an interface to the R
statistical package. Although not a web application, it is based on the
Qt webkit framework and uses web based technology in the background.
R-Studio came into the picture when the COMMIT/ project provided a grant
for developing SWISH as a toolkit for analysis of relational (SQL) data.
The R-studio interface has a similar layout as SWISH, providing a source
window, a console and an output plane that typically shows results in
tables or charts.

\item
IPython Notebook \cite{rossant2013learning} allows mixing markdown or
HTML text with Python sources. The rendered Notebook shows the text,
sources and possible results in the form of numbers, tables or charts.
\end{itemize}

SWISH embodies most of the ideas behind JSFiddle and R-Studio. Embedding
of SWISH in documents is demonstrated in \secref{lpn}. Interactive
editing of documents that embed SWISH is discussed in future work
(\secref{future}).

Both R-Studio and IPython Notebook work on the basis of
\jargon{authentication} (either to the OS or application), after which
any command may be executed.  SWISH can operate both as a public service
granting access to non-intrusive queries and as an authenticated service
to run arbitrary queries, for example for maintenance purposes.  See
\secref{cliopatria}.

\section{The SWISH application}
\label{sec:application}

SWISH consists of two parts. The \emph{client side}, running in a
browser, is implemented as a series of jQuery plugins, using Bootstrap
for styling and RequireJS\footnote{\url{http://requirejs.org/}} for
package management. The \emph{server side} is completely implemented in
SWI-Prolog \cite{DBLP:journals/tplp/WielemakerHM08}. It builds on top of
the SWI-Prolog HTTP server libraries, the Pengines library and the IDE
support libraries that provide data for auto completion, documentation
and highlighting.

In the following sections we describe SWISH in terms of interface
components, where we discuss the requirements, the user aspects, the
client code and supporting server functionality	for each component.
First, we provide a screendump that illustrates the main components
in \figref{swish}.

\begin{figure}
    \includegraphics[width=\linewidth]{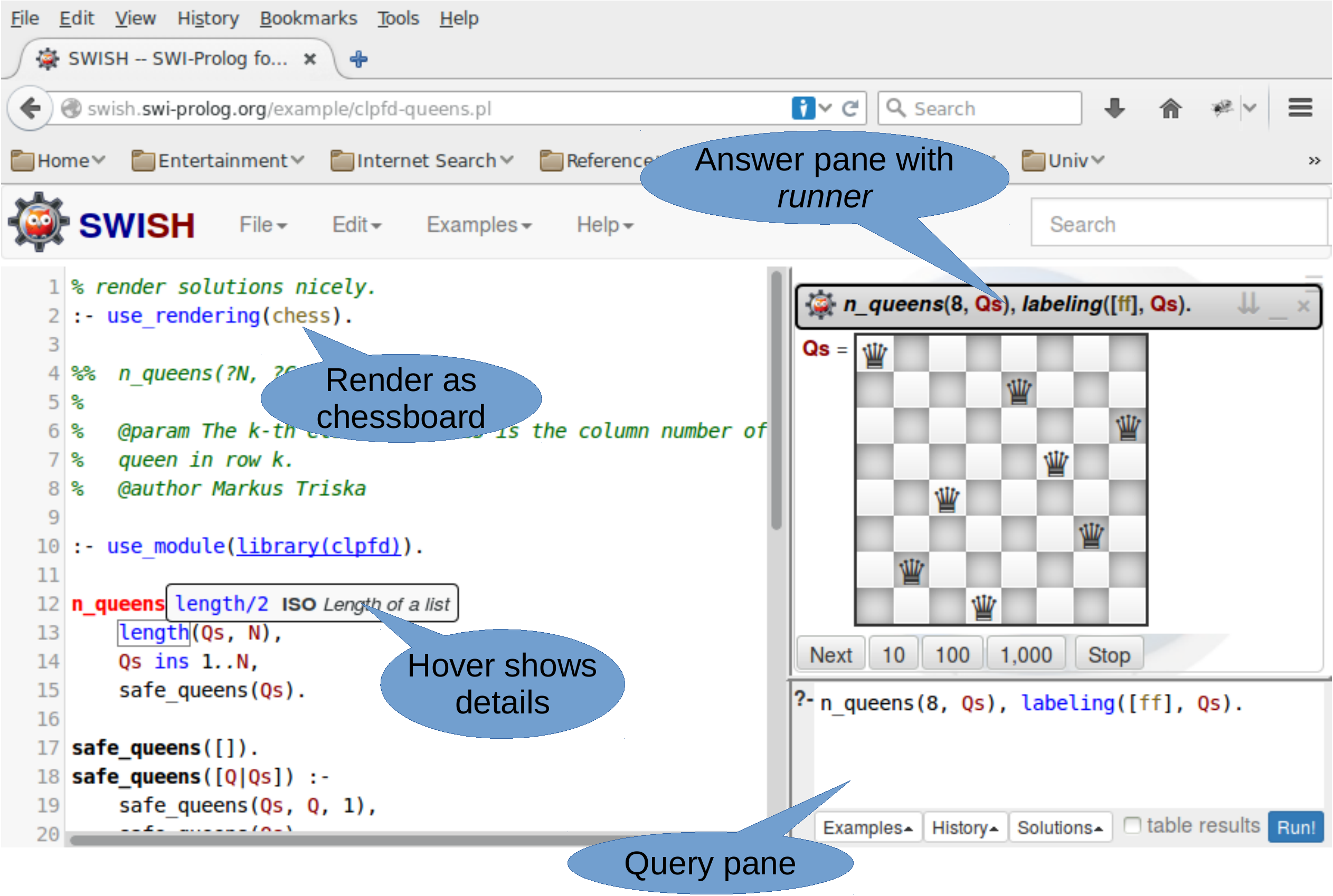}
    \caption{Screendump of SWISH. The left pane shows the source code,
	     while the top-right pane shows a query \jargon{runner} that
	     exploits the current selected answer \jargon{renderer} and
	     buttons on how to continue after the first answer. The
	     bottom-right pane provides the query editor with access to
	     example queries stored in the source, query history, apply
	     solution modifiers, result presentation and a \textsf{Run!}
	     button to start the query.}
    \label{fig:swish}
\end{figure}

\subsection{The code editor}

A proper editor is the most important component of a usable programming
environment. The editor must support the language, including syntax
highlighting, auto indentation, code completion based on templates and
already existing code and highlighting of errors and warning from the
compiler.  The editor is used both for editing the source code and
editing queries.

Prolog is a difficult language to support in code editors due to the
lack of reserved keywords, e.g., the word \const{if} in C starts an
if-statement if not embedded in comment or a string, but the word
\const{is} in Prolog can refer to the built-in predicate \index{is/2}\predref{is}{2}, but also
some predicate with a different arity, just a constant, etc. Another
example is \exam{X-Y} which can both be an arithmetic expression or a
\jargon{pair} as used with e.g., \index{keysort/2}\predref{keysort}{2}. Next to the lack of keywords
the ability to extend the syntax using new operators complicates the
implementation of syntax support while editing. SWI-Prolog's built-in
Emacs oriented editor resolves this problem by closely integrating
Prolog with the editor. While typing, the current term (clause or
directive) is parsed and analysed in the context of the current file and
the file's imports after each keystroke. If the term has valid syntax,
all tokens are coloured according to their syntactic role as well their
relation to the remainder of the program. For example, a call to a
non-existing predicate is coloured red, a call to a built-in or imported
predicate is blue and a call to a locally defined predicate is black.
The libraries that implement this analysis have been decoupled from the
built-in editor, both to support source colouring for the SWI-Prolog
documentation system PlDoc \cite{Wielemaker:2007c} and
ProDT\footnote{\url{http://prodevtools.sourceforge.net}, these libraries
are not yet used by ProDT.}

There are two dominant open source and actively maintained in-browser
code editors available: ACE and CodeMirror. When we started SWISH, ACE
had a very basic Prolog mode and CodeMirror had none. We nevertheless
opted for CodeMirror because its highlighting is based on raw JavaScript
code rather than a regular expression based template language as used
for ACE. The low level implementation allows for a novel highlighting
implementation. The highlighter consists of a JavaScript implemented
Prolog \emph{tokeniser}. Tokenizing Prolog is sufficient to colour
comments, quoted material (strings, quoted atoms), variables and
constants (atoms and numbers). It is also sufficient to support smart
indentation. As discussed above, it is not sufficient for highlighting
the role played by atoms and compound terms.\footnote{An additional
complication is formed by CodeMirror's token-based highlighting which
does not support look-ahead. As a consequence, we must decide on the
colour of e.g., \exam{asserta(} while we do not know the arity of the
term.}

\begin{figure}
    \includegraphics[width=\linewidth]{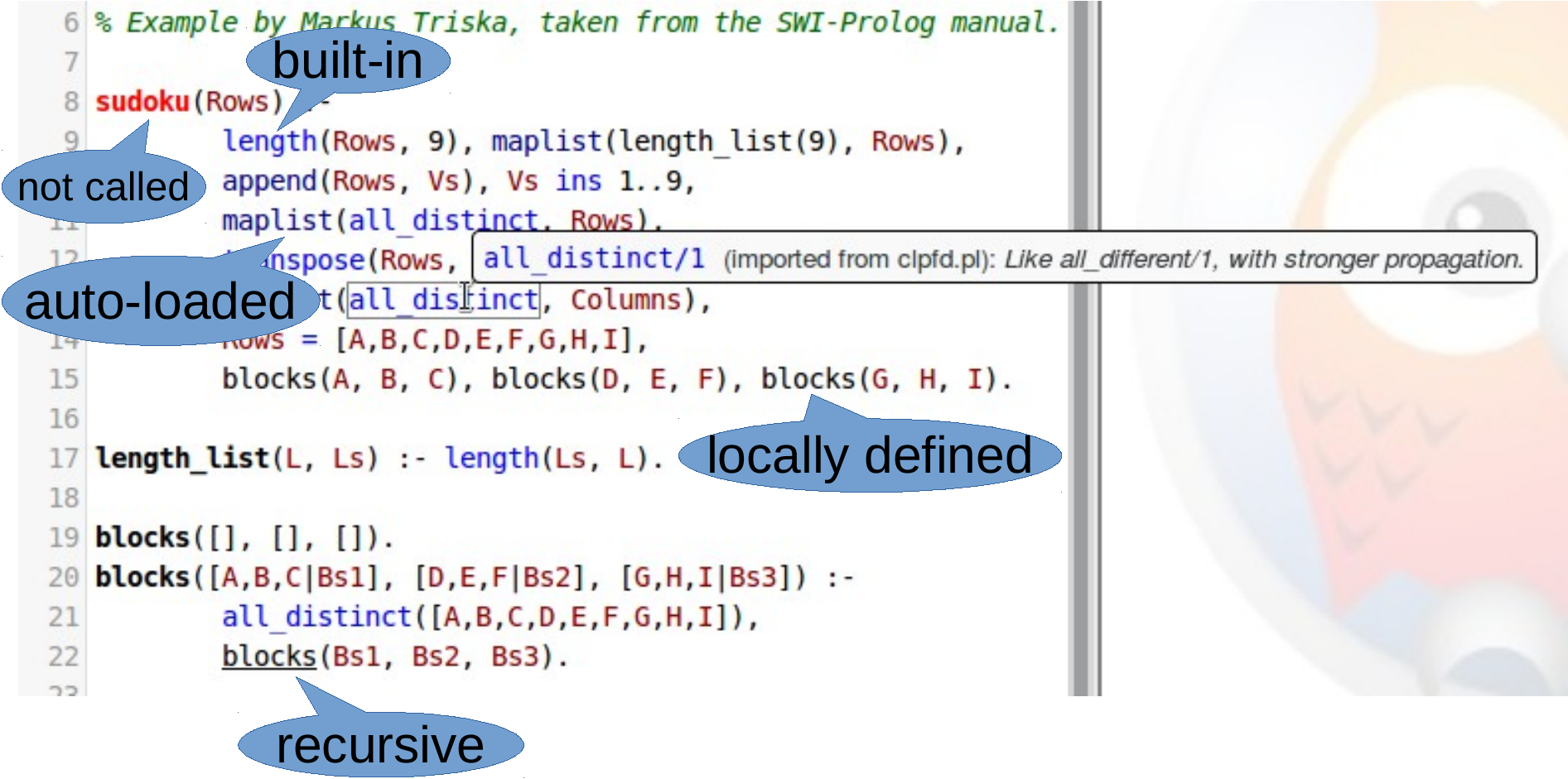}
    \caption{The semantic highlighter classifies, in addition to
	     the syntactic category such as \jargon{comment} or
	     \jargon{variable}, terms that define or call predicates
	     based on cross-referencing the source code.}
    \label{fig:semhighlight}
\end{figure}

We provide \jargon{semantic} highlighting as illustrated in
\figref{semhighlight} by (1) forwarding the changes to the content of
the editor to the server server which maintains a mirror of the editor
content and (2) asking the server to produce a list of semantically
enriched tokens for the source. The tokens are returned as a
list-of-lists, where each inner list represents the tokens for a source
term (clause or directive). Grouping the tokens per source term allows
for incremental update (not yet implemented) as well as
re-synchronisation (see below). For example, a fragment of an enriched
token list may look like this: [ \textit{functor, (undefined_goal)},
\textit{variable (singleton)}, \ldots ]. The JavaScript tokeniser
matches its tokens with this list. If the basic type (e.g., `functor' or
`variable') matches, it uses the enrichment information (e.g.,
`singleton') to decide on the style. If the basic token type does not
match, it highlights the token using the basic syntactical
category and schedules a request to the server for a new list of
enriched tokens. This request is sent if the user pauses typing for 2
seconds. The request is accompanied by the full source if this is small
or the list of changes since the last request if the source is large.
While waiting for up-to-date enriched tokens, the JavaScript
highlighting code heuristically tries to re-synchronise and either uses
the uncertain results or falls back to the plain tokens.
Re-synchronisation checks whether a single added, deleted or modified
token gets the token stream in-sync. If this fails it tries to
re-synchronise on a full-stop with the next clause or directive.

A CodeMirror \jargon{hover} plugin is used to show basic information
about tokens if the pointer hovers over it. For goals, this includes
where the goal is defined (ISO, SWI-Prolog built-in, a library, locally)
and the documentation summary information if available. This information
is requested from the server.

A CodeMirror \jargon{template} plugin is configured from templates
(e.g., \exam{atom_length(+Atom, -Length)}) extracted from the SWI-Prolog
manual and PlDoc documentation of imported libraries. This plugin shows
a menu of applicable predicates with their templates on
\textsf{Control-Space}.

Finally, if the user uses the \textsf{Run!} button to execute a query,
the program is compiled. If the compiler generates errors or warnings,
these are inserted as notes in the source code.

\subsection{Source code and query management}
\label{sec:gitty}

As JSFiddle formed the initial inspiration for SWISH, SWISH has a
facility to save the program. The current version of SWISH explicitly
targets the cooperative development of Prolog programs and queries
related to a dataset (see \secref{cliopatria}). This triggered the
implementation of a more organised storage facility. The server-side
storage module is implemented in Prolog and inspired by GIT. Each file
is versioned independently rather than maintaining the version of a
hierarchy of files. Files can be referenced by content using their GIT
compatible SHA1 hash or by name. The name can be considered a
version \jargon{head} and refers to the latest version with that name.
The file save and load interface provides the following operations:

\begin{itemize}
    \item Saving a file anonymously, which produces a randomly
          generated URL similar to JSFiddle.
    \item Saving a file by name.
    \item Saving a new version.  The interface shows the available
          versions and the modifications.
    \item Forking a file under a new name.  The history remains
          linked to the original.
\end{itemize}

Prolog source files can \jargon{include} other sources on the same
server using \exam{:-~include(filename).}, including the latest
version or \exam{:-~include(hash).} to include a specific version.

Prolog source files can embed \jargon{example queries} using structured
comments, where each sequence from \verb$?-$ to the matching full stop
token is added to the \textsf{Examples} menu of the query panel (see
\figref{lpn}). The comment below illustrates a call to \index{append/3}\predref{append}{3} embedded
in the source window.

\begin{code}
/** <examples>

?- append([one], [two,three], List).
*/
\end{code}

\noindent
\subsection{The query editor}
\label{sec:queryeditor}

The \jargon{query editor} is based on the same jQuery plugin that
realises the code editor and thus profits from the syntax
highlighting, template insertion and hovering plugins.  In addition,
it provides three popup menus:

\begin{description}
    \item[Examples] This menu is filled from the structured comments
described above.  The examples menu is shown in \figref{lpn}.
    \item[History] This menu provides earlier executed queries.
    \item[Solutions]  This menu embeds an existing query in a meta-call to
alter the result. Currently provided operations are \textit{Aggregate
(count all)}, \textit{Order by}, \textit{Distinct}, \textit{Limit},
\textit{Time} and \textit{Debug (trace)}.  \Figref{aggregate} shows
how the menu is used to count the solutions of a goal.
\end{description}

\begin{figure}
    \includegraphics[width=\linewidth]{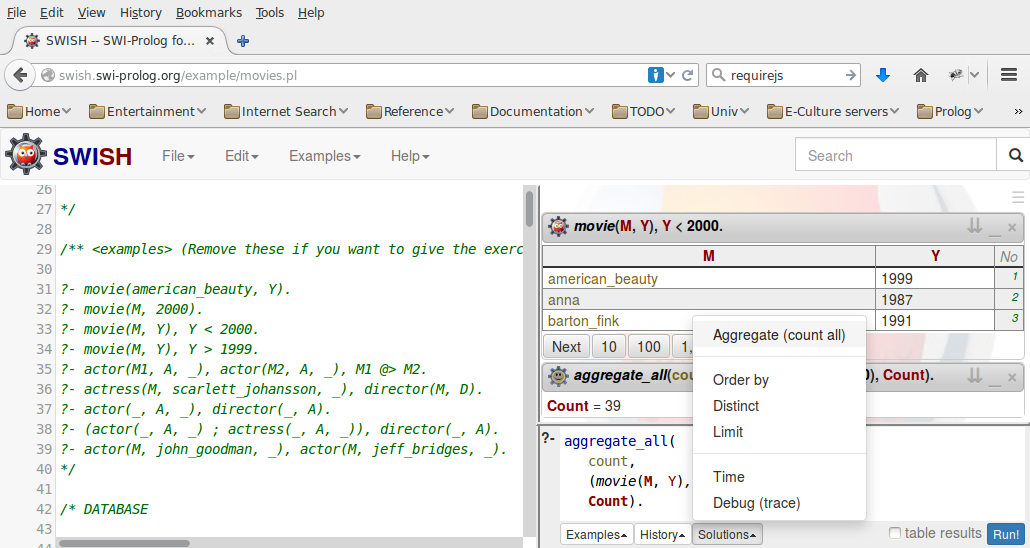}
    \caption{The \textsf{Solutions} menu can be used to count results,
	     order them, filter duplicates, etc. The upper runner shows
	     answers to the query as a table.}
    \label{fig:aggregate}
\end{figure}

\subsection{Running a query: runners in the answer pane}
\label{sec:answerpane}

The answer pane is a placeholder for \jargon{runners}, where each runner
represents a query. The answer pane provides a menu for operations on
all runners inside it. Provided actions are \textsf{Collapse all},
\textsf{Expand all}, \textsf{Stop all} and \textsf{Clear}. The query may
be completed, running or waiting for user input. SWISH can manage
multiple active queries at the same time, up to an application defined
maximum (default~3).

Each runner provides its own set of commands to control the specific
query. During execution a runner provides an \textsf{Abort} button.
After query evaluation completes with an answer and more answers may be
available the runner allows for asking the next 1, 10, 100 or 1,000
results or to \textsf{Stop} the query. In addition, the runner shows a
\jargon{text input} field when the application wants to read input and
may show debugger interaction buttons if the tracer is being used (see
\secref{tracer}).

A runner can render answers in two modes, the classical Prolog mode or
as a table, similar to what many database interfaces provide. The
`table' mode is particularly appealing for querying datasets (see
\figref{aggregate}), while the former is more suitable for rendering
small amounts of complex answers such as the chessboard position in
\figref{swish}. By default, Prolog terms are rendered as structured HTML
objects, where the rendered text is the same as Prolog's \index{writeq/1}\predref{writeq}{1}
predicate.

The server can provide \jargon{rendering libraries} that render Prolog
terms using dedicated HTML. In \figref{swish}, the `chess' renderer is
loaded due to the \exam{:- use_rendering(chess)} directive. The `chess'
renderer translates a list of length $N$ holding integers in the range
$1..N$ as a chessboard with queens. In addition to the chess rendering
library, SWISH provides rendering libraries for sudoku puzzles, parse
trees and tables. The ClioPatria version adds a renderer for RDF
resources that renders the resource more compactly and provides a
hyperlink for obtaining details. If a term can be rendered in multiple
ways, the interface provides a hover menu to select between the
alternatives. \Figref{render} illustrates this functionality. The render
facility is similar to the Prolog \index{portray/1}\predref{portray}{1} hook that allows changing
the result printed for terms with a specific shape. However, it can
exploit the full potential of HTML (or SVG) and the interface allow for
switching the selected rendering.

A rendering library is a module that must define a non-terminal (grammar
rule) \index{term_rendering//3}\dcgref{term_rendering}{3}, calling
\index{html//1}\dcgref{html}{1}\footnote{http://www.swi-prolog.org/pldoc/doc_for?object=html/3}
to produce HTML from the Prolog input term, a list of variable bindings
(Name~=~Variable) and user provided options. In the current version, new
rendering modules must be loaded into the SWISH server and cannot be
created by the SWISH user.

\begin{figure}
    \includegraphics[width=\linewidth]{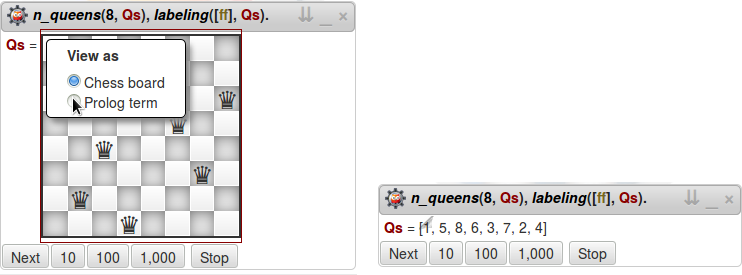}
    \caption{With the `chess' render library, a list of integers is
	     interpreted as queens on a chessboard.  The user can
	     select rendering as a `Prolog term' to see the actual
	     term.}
    \label{fig:render}
\end{figure}

\subsubsection{Server side execution of the query}
\label{sec:pengines}

Server-side execution of a query is supported by the Pengines
\cite{DBLP:journals/tplp/LagerW14} library. The Pengines library allows
for creating a Prolog engine represented by a Prolog thread. Optionally,
the pengine is handed a Prolog program that is loaded into the pengine's
workspace (program space). The workspace is a temporary module that is
disposed of after the pengine terminates. The pengine may be asked
questions through HTTP queries, similar to a traditional Prolog user
interacting with Prolog running in a
terminal.\footnote{\url{https://www.youtube.com/watch?v=G_eYTctGZw8}}

If the SWISH user hits the \textsf{Run!} button, the content of the
source pane is used to create a new pengine. Subsequently, the content
of the query pane is sent as the one and only query that will be
executed by the pengine.\footnote{Pengines can execute multiple queries.
We do not use this facility because a fresh pengine starts in a
predictable state (standard operators, empty dynamic database).} Before
execution, the query is verified to be \emph{safe}, unless sandboxing is
disabled (see \secref{cliopatria}). The sandbox component is discussed
below.

The pengine's default working module may be pre-loaded with code. SWISH
uses this facility to redefine the Prolog I/O predicates such as \index{read/1}\predref{read}{1},
\index{write/1}\predref{write}{1}, \predref{format}{1,2,3}, etc. The ClioPatria version
(\secref{cliopatria}) also preloads the RDF libraries, so users can run
queries on the RDF database without explicitly importing the required
libraries.

\subsubsection{Sandboxing queries}
\label{sec:sandbox}

A Prolog environment contains global state in the form of loaded
modules, defined operators, dynamic predicates, etc. Prolog exposes a
rich and potentially dangerous interface to the operating system. For an
anonymous services, we want each query to start in a well defined state
and we must ensure that execution of the query does not make unwanted
changes to the hosting computer or leaks sensitive information.

Both for education purposes and data analysis one can write meaningful
programs without making permanent changes to the server or the server's
filesystem. That is where the sandbox library comes in. The sandbox
library is active while loading the source, where it refuses to add
clauses to other modules than the pengine's workspace and where it only
accepts a restricted set of \jargon{directives}, also aimed at keeping
all changes local to the workspace. Prior to execution, the sandbox
unfolds the query and compares all reachable goals with a whitelist. The
whitelist contains all side-effect free built-in Prolog predicates, safe
meta-predicates (e.g., \index{findall/3}\predref{findall}{3}) and allows for using the dynamic
database, provided that the head of the affected predicate is not
module-qualified (and thus the referenced predicate lives in the
temporary program space of the Pengine) and the body is safe. It does
\emph{not} allow for cross-module calls (Module:Goal) to
\jargon{private} predicates and does not provide access to
object-enumeration predicates such as \index{current_atom/1}\predref{current_atom}{1},
\index{current_predicate/1}\predref{current_predicate}{1}, etc., both to avoid leaking sensitive information.

The sandbox test fails under one of these conditions:

\begin{itemize}
    \item It discovers a (meta-) goal for which it cannot deduce the
    called code.  The traditional example is \exam{read(X), call(X)}.
    If such a goal is encountered, it signals an \jargon{instantiation
    error}, together with a trace that explains how the insufficiently
    instantiated goal can be reached. Note that it can deal with normal
    high-order predicates if the meta-argument is specified. For
    example, the following goal is accepted as safe.

    \begin{code}
    ?- maplist(plus(1), [1,2,3])
    \end{code}

\noindent
    \item It discovers a goal that is not whitelisted.  In this case
    it signals a \jargon{permission error}, again accompanied with
    a trace that explains how the goal can be reached.  Note that
    pure Prolog predicates are unfolded, also if it concerns predicates
    from the libraries or belonging to the set of built in predicates.

    \item It discovers a cross-module (\arg{M:Goal}) call to a predicate
    that is not public. Normally, SWI-Prolog, in the tradition of
    Quintus Prolog, allows for this. Allowing it in SWISH would imply
    that no data can be kept secret. With this limitation, libraries can
    keep data in local dynamic predicates that remain invisible to
    non-authorised users.
\end{itemize}

\subsubsection{Debugging}
\label{sec:tracer}

The SWISH debugger is based on the traditional 4-port debugging model
for Prolog. \Figref{tracer} shows the tracer in action on \index{sublist/2}\predref{sublist}{2} from
the \textit{Lists} example source. The debugger was triggered by a
break-point on line~10 set by clicking on the line-number in the code
editor. The debugging interaction is deliberately kept simple and
similar to traditional programming environments. A \jargon{retry} button
is added to the commonly seen `step into', `step over' and `step out'
for highlighting the unique feature of Prolog to re-evaluate a goal.

\begin{figure}
    \includegraphics[width=\linewidth]{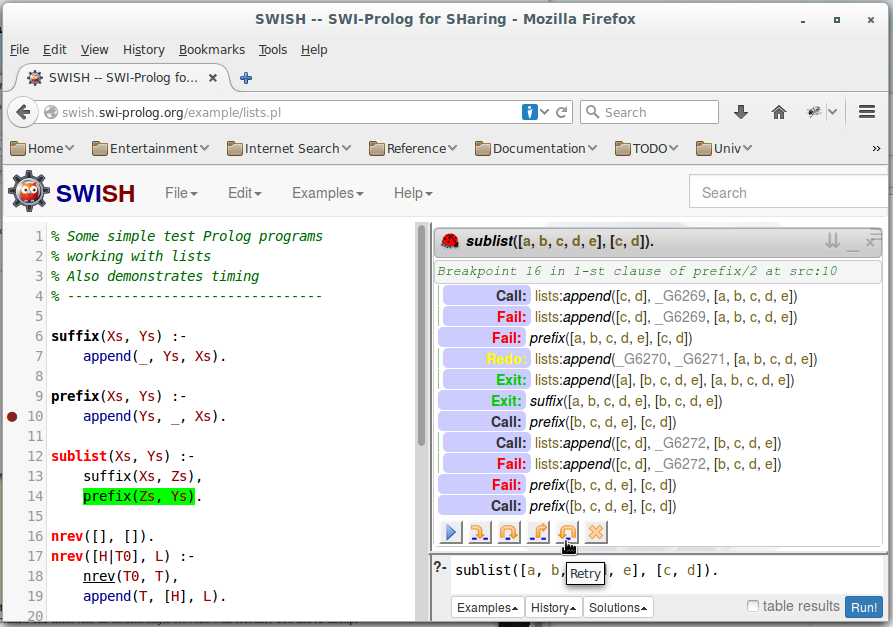}
    \caption{Debugging applications in SWISH}
    \label{fig:tracer}
\end{figure}

\section{Portability}
\label{sec:portability}

The SWISH client libraries are based on mature and well maintained
JavaScript libraries. The client runs all modern major browsers with
HTML5, CSS and JavaScript support. It is frequently tested on FireFox,
Chrome, Safari and Internet Explorer~11.

The server code is basically non-portable. Many of the required
libraries and features are shared with at least one other Prolog
implementation, but none is capable to support the full range. Below we
summarise the main problems.

\begin{itemize}
    \item The scale of the involved Prolog libraries demands for
    a closely compatible Prolog module system.  Probably only
    SICStus and YAP can be used without significant rewrites.

    \item The HTTP server libraries are heavily based on C~code
    that interacts with the SWI-Prolog foreign language interface
    to Prolog streams.  YAP has copied the low-level libraries and is
    capable to run (an old version of) these libraries.

    \item The Pengines library depends on the HTTP library and
    the multi-thread interface.  The SWI-Prolog thread API is
    also provided by YAP and XSB.

    \item The sandbox library (\secref{sandbox}) assumes that
    whitelisted predicates are indeed safe.  This requires robust
    handling of invalid calls and resource overflows.  Few Prolog
    systems can satisfy this requirement.  SICStus Prolog would
    be a candidate, but SICStus does not support multi-threading.

    \item The semantic syntax highlighting depends on detailed
    source layout information provided by \index{read_term/3}\predref{read_term}{3}.  SWI-Prolog's
    support for term layout is an extended version of the Quintus
    Prolog term layout functionality.

    \item Significant parts of the code rely on SWI-Prolog version~7
    extensions, notably the \jargon{dict} and \jargon{string} types
    that facilitate a natural mapping between Prolog and JSON data.
\end{itemize}

From the above list it should be clear that a fully functional port of
SWISH to another Prolog system is not immediately feasible. YAP probably
comes closest but still requires a significant amount of work.

There is a more realistic scenario though. In this setup, SWI-Prolog
provides the web interface and most of the development tools and another
language, not even necessarily Prolog, provides the query solving. The
interface between the two can be based on interprocess communication or,
if the target system is robust, safe and capable of supporting
threads, by linking the target system into the process and using the
C~interface for communication.

\section{Applications}
\label{sec:applications}

In this section we describe and evaluate four publicly available SWISH
applications. All these services are regularly updated to run the latest
version of SWISH and SWI-Prolog.

\subsection{SWISH}
\label{sec:swishapp}

SWISH\footnote{\url{http://swish.swi-prolog.org}} runs a plain publicly
accessible copy of SWISH that allows running \jargon{sandboxed} (see
\secref{sandbox}) Prolog programs. The server has collected 10,800
programs between September 29, 2014 and June 2, 2015. Over the month May
2015, it has executed 258,809 Prolog queries. The web site is regularly
used by users of the \#\#prolog IRC channel to discuss programming
solutions and is in active use for education.\footnote{Steve Matuszek,
UMBC (via e-mail: ``Thank you very much for this fantastic resource! I
used it while teaching Prolog this semester, and it really helped
tighten the loop for my students. We spent zero time on tool
installation and other overhead, and all the time on understanding the
concepts. I even had them turn their assignments in via SWISH, with
their test queries in the examples block.''}

\subsection{ClioPatria}
\label{sec:cliopatria}

ClioPatria is a semantic web (RDF) framework for SWI-Prolog. It consists
of an RDF triple store, a SPARQL server and a web frontend to manage the
server and explore the data in the RDF store. ClioPatria can be extended
using \jargon{cpacks} (ClioPatria pack or plugin). SWISH is available
as a ClioPatria
cpack\footnote{\url{http://cliopatria.swi-prolog.org/packs/swish}}
and makes the Prolog shell available for querying as
well as maintenance tasks. Without login, user can run side-effect free
queries over the RDF data stored in ClioPatria's RDF database. After
login with administrative rights, the sandbox limitations are lifted and
the Prolog shell can be used to perform maintenance tasks on the RDF
data such as data transformation, cleanup, etc., as well as program
maintenance tasks such as reloading modified source files.

SWISH has been used in the Talk Of Europe creative
camp\footnote{\url{http://www.talkofeurope.eu/}} to explore data on the
speeches in the European
parliament.\footnote{\url{http://purl.org/linkedpolitics}} Although
still immature, users appreciated the ability to define more efficient
and expressive queries than provided by the SPARQL query interface.
Above all, the ability to save and share programs that perform
interesting tasks on the data was frequently used, in particular to seek
help fixing queries.

\subsection{Learn Prolog Now!}
\label{sec:lpn}

Learn Prolog Now!\footnote{\url{http://www.learnprolognow.org}} is an
online version of a Prolog book by Patrick Blackburn, Johan Bos, and
Kristina Striegnitz \cite{blackburn2006learn}. We established a proof of
concept that embeds SWISH in the online course
material.\footnote{\url{http://lpn.swi-prolog.org}} It is realised as a
Prolog hosted \jargon{proxy} that fetches the pages from the main site
and serves the enhanced pages to the user. The proxy identifies and
classifies the code fragments in terms of `source code' `queries' and
dependencies. Next, it adds a button to the source fragments that, when
pressed, replaces the HTML \verb$<pre>$ element with in \verb$<iframe>$
running SWISH filled with the source while the example queries are added
to the \textsf{Examples} menu (\figref{lpn}). The queries are executed
by \url{http://swish.swi-prolog.org}. Program and queries are
transferred using the following HTTP parameters: \const{code} (the
source code), \const{background} (source code that is loaded into the
pengine but not visible in the editor), \const{examples} (queries that
appear in the \textsf{Examples} menu) and \const{q} (the initial query).
The proxy server served 19,700 pages during May 2015.

\begin{figure}
    \includegraphics[width=\linewidth]{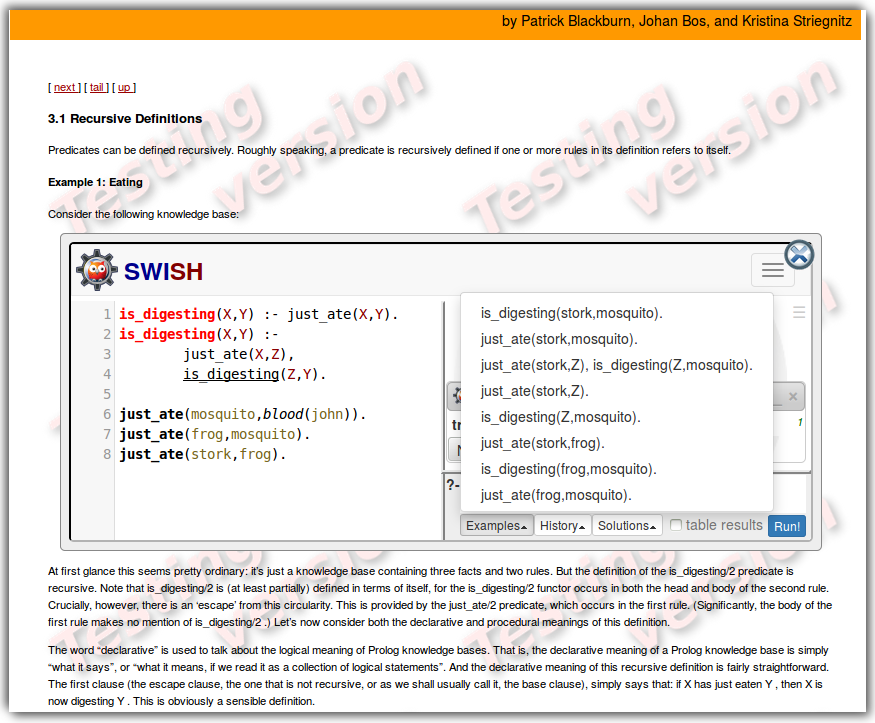}
    \caption{Screendump of Learn Prolog Now with opened SWISH instance
	     that shows the collected source as well as example queries
	     from the subsequent text that are classified as relating to
	     this source.  The embedded SWISH provides all functionality
	     available in the stand-alone SWISH. If the user presses the
	     close button, SWISH will be removed and the original code
	     re-appears.}
    \label{fig:lpn}
\end{figure}

\subsection{\texttt{cplint} on SWISH}
\label{sec:cplint}

\texttt{cplint} on SWISH\footnote{\url{http://cplint.lamping.unife.it/}}
is a web application based on SWISH for reasoning with probabilistic logic programs under the distribution semantics. The Prolog source window
is used to write a logic program with annotated disjunction. A query in the form of a ground atom is answered by returning its probability of being true in the program. The computation of the probability is done with the \texttt{cplint} system \cite{RigSwi13-TPLP-IJ}  in the server using Pengines. The input program is translated into an internal representation using source to source transformation.

 SWISH was modified only in the JavaScript code for the runner. The source code is prepended code for loading the  \texttt{cplint} library and enabling the source to source transformation while the query is wrapped into a call of the inference predicate. This call has a variable argument \verb|Prob| which will hold the probability and will be shown to the user in the answer pane.

\subsection{TRILL on SWISH}
\label{sec:trill}

TRILL on SWISH\footnote{\url{http://trill.lamping.unife.it/}} is a
probabilistic OWL reasoner that reuses SWISH. As SWISH for ClioPatria,
described in \secref{cliopatria}, it is a ClioPatria \jargon{cpack}. The
Prolog source window is replaced by an RDF/XML editor window that can be
used to upload an OWL ontology while the query editor can be used to
pose Prolog queries against the OWL ontology. The probability of queries
is computed using TRILL \cite{RigBel14-URSWb-BC}, a reasoner in Prolog
that adopts the distribution semantics for probabilistic description
logics.

Also for TRILL on SWISH we had only to modify the JavaScript code for the runner. The source code sent to the Pengine is obtained by adding Prolog code for parsing an RDF/XML string,  by calling the parsing predicate and by wrapping the query in a meta-call that performs syntactic checks for misspellings.

\section{Future work}
\label{sec:future}

Although definitely usable in its current state, SWISH is work in
progress. We are confident that the basic component selection and
organisation of the server and client code are stable. More work is
needed to improve the current system. Notably the semantic highlighting
is not yet perfect and often fails to degrade gradually if the server
side annotation does not match the client tokens perfectly. The
Pengine's sandbox protection is often too restrictive, while several
security flaws have been reported and fixed already. It is, and probably
always will be, advised to run public SWISH-enabled services in an
operating system sandbox. The current server suffers from memory leaks
and stability problems. Although the situation has improved
significantly, the main demo server needs to be restarted about once a
week.\footnote{A restart of the server has only small consequences to
active users. Open queries are killed. The source code mirror is lost,
but automatically recovered if the client asks for a new set of
semantically enriched tokens.}

We foresee several extensions to SWISH that will improve current
applications and enable new opportunities for deploying SWISH.

\begin{itemize}
    \item ClioPatria's authorised usage of SWISH shows some of
the potential for controlling servers or embedded Prolog engines.
In addition to small temporary Prolog programs, we would like to
be able to edit existing and create new Prolog modules as well as
pages in other languages, such as JavaScript, HTML and CSS.  Full
editing capabilities would allow for shared development of server
software without shell access to the server on which SWISH enabled
software is running.

    \item Multi-document editing can enhance the sandboxed SWISH
application by providing input and output documents.  Compare this
to TRILL (\secref{trill}) using an RDF/XML document as input.

    \item We plan to provide a markdown-based format specifically
for writing tutorials and well as dataset analysis documents. The first
will look like the Learn Prolog Now! example discussed in \secref{lpn}.
For the second, we envision a document with embedded code and query
fragments, where the query fragments produce tables or charts.  This
is similar to IPython Notebook.

    \item Turn \url{http://swish.swi-prolog.org} into a reliable
and scalable resource. Examine the possibility for schools to
instantiate a private version that is preloaded with course material and
assignments.
\end{itemize}

\section{Conclusion}
\label{sec:conclusions}

This article presented SWISH, \textbf{SWI}-Prolog for \textbf{Sh}aring.
SWISH provides a web-enabled interface to Prolog that is based on ideas
from JSFiddle, R-Studio and IPython Notebook. It consists of a
JavaScript client side, while the server side is based on SWI-Prolog's
HTTP and Pengines (Prolog engines) libraries. SWISH can be deployed in
many settings, such as education, data analysis and server development
and maintenance. SWISH as a whole is tied to SWI-Prolog, but other
languages, not even limited to Prolog, could be controlled from
SWI-Prolog. SWISH is made available as open source and can be downloaded
from github.\footnote{\url{https://github.com/SWI-Prolog/swish}}

\subsection*{Acknowledgements}

The development of SWISH was supported by the Dutch national program
COMMIT/.

\bibliographystyle{plain}
\bibliography{iulp-1}

\begin{thebibliography}{1}

\bibitem{blackburn2006learn}
Patrick Blackburn, Johan Bos, and Kristina Striegnitz.
\newblock {\em Learn prolog now!}, volume~7.
\newblock College Publications, 2006.

\bibitem{gandrud2013reproducible}
Christopher Gandrud.
\newblock {\em Reproducible Research with R and R Studio}.
\newblock CRC Press, 2013.

\bibitem{DBLP:journals/tplp/LagerW14}
Torbj{\"{o}}rn Lager and Jan Wielemaker.
\newblock Pengines: Web logic programming made easy.
\newblock {\em {TPLP}}, 14(4-5):539--552, 2014.

\bibitem{RigSwi13-TPLP-IJ}
Fabrizio Riguzzi and Terrance Swift.
\newblock Well\--definedness and efficient inference for probabilistic logic
  programming under the distribution semantics.
\newblock {\em Theory Pract. Log. Program.}, 13(Special Issue 02 - 25th Annual
  GULP Conference):279--302, March 2013.

\bibitem{rossant2013learning}
Cyrille Rossant.
\newblock {\em Learning IPython for interactive computing and data
  visualization}.
\newblock Packt Publishing Ltd, 2013.

\bibitem{Wielemaker:2007c}
Jan Wielemaker and Anjo Anjewierden.
\newblock {PlDoc}: {Wiki} style literate programming for {Prolog}.
\newblock In Patricia Hill and Wim Vanhoof, editors, {\em Proceedings of the
  17th Workshop on Logic-Based methods in Programming Environments}, pages
  16--30, 2007.

\bibitem{DBLP:journals/tplp/WielemakerHM08}
Jan Wielemaker, Zhisheng Huang, and Lourens van~der Meij.
\newblock Swi-prolog and the web.
\newblock {\em {TPLP}}, 8(3):363--392, 2008.

\bibitem{RigBel14-URSWb-BC}
Riccardo Zese, Elena Bellodi, Evelina Lamma, Fabrizio Riguzzi, and Fabiano
  Aguiari.
\newblock Semantics and inference for probabilistic description logics.
\newblock In {\em Uncertainty Reasoning for the Semantic Web III}, volume 8816
  of {\em LNCS}, pages 79--99. Springer, 2014.

\end{thebibliography}

\egroup

\end{document}